\documentclass[conference,a4paper]{IEEEtran}
\usepackage{cite}
\usepackage{graphicx,color,epsfig,rotating}
\usepackage{amsfonts,amsmath,amssymb,bbm}
\usepackage{algorithm}% http://ctan.org/pkg/algorithms
\usepackage{algpseudocode}% http://ctan.org/pkg/algorithmicx
\usepackage{subfigure}
\usepackage{amsmath}
\usepackage{cite}
\usepackage{mdwtab}
\usepackage{placeins}
\usepackage{psfrag, graphicx}
\usepackage[latin1]{inputenc}
\usepackage{amssymb}
\usepackage{multirow}
\usepackage{stfloats}
\usepackage{tabularx} 
\usepackage{booktabs} 
\usepackage{url}

\long\def\comment#1{}

\newfont{\bbb}{msbm10 scaled 700}

\newfont{\bb}{msbm10 scaled 1100}

\newcommand{\Wc}{{\cal W}}

\newcommand{\be}{\begin{equation}}
\newcommand{\ee}{\end{equation}}
\newcommand{\bea}{\begin{eqnarray}}
\newcommand{\eea}{\end{eqnarray}}

\newtheorem{defn}{Definition}%[section]
\newtheorem{example}{Example}%[section]
\newtheorem{theorem}{Theorem}%[section]
\makeatletter
\newcommand{\subalign}[1]{%
  \vcenter{%
    \Let@ \restore@math@cr \default@tag
    \baselineskip\fontdimen10 \scriptfont\tw@
    \advance\baselineskip\fontdimen12 \scriptfont\tw@
    \lineskip\thr@@\fontdimen8 \scriptfont\thr@@
    \lineskiplimit\lineskip
    \ialign{\hfil$\m@th\scriptstyle##$&$\m@th\scriptstyle{}##$\crcr
      #1\crcr
    }%
  }
}
\makeatother

\ifodd 1

\else

\fi

\ifodd 1

\else

\fi

\begin{document}
\sloppy
\title{Cache-aided Interference Management Using Hypercube Combinatorial Cache Designs}
\author{
\IEEEauthorblockN{Xiang Zhang, Nicholas Woolsey, Mingyue Ji }
	\IEEEauthorblockA{Department of Electrical and Computer Engineering, University of Utah\\
		Salt Lake City, UT, USA\\
		Email: \{xiang.zhang, nicholas.woolsey, mingyue.ji\}@utah.edu}}
\maketitle

\begin{abstract}
We consider a cache-aided interference network which consists of a library of $N$ files, $K_T$ transmitters and $K_R$ receivers (users), each equipped with a local cache of size $M_T$ and $M_R$ files respectively, and connected via a discrete-time additive white Gaussian noise channel. Each receiver requests an arbitrary file from the library. The objective is to design a cache placement without knowing the receivers' requests and a communication scheme such that the sum Degrees of Freedom (sum-DoF) of the delivery is maximized. This network model has been investigated by Naderializadeh {\em et al.}, who proposed a prefetching and a delivery scheme that achieve a sum-DoF of $\min\{\frac{{M_TK_T+K_RM_R}}{{N}}, K_R\}$. One of the biggest limitations of this scheme is the requirement of high subpacketization level. {This paper attempts to design new algorithms to reduce the file subpacketization in such a network. }%This paper is the first attempt in the literature (according to our knowledge) to reduce the file subpacketization in such a network. 
In particular, we propose a new approach for both prefetching and linear delivery based on a combinatorial design called {\em hypercube}. We show that the required number of packets per file can be exponentially reduced compared to the state-of-the-art scheme proposed by Naderializadeh {\em et al.}, or the NMA scheme. When $\frac{M_TK_T+K_RM_R}{N} \leq K_R$, the achievable one-shot sum-DoF using this approach is $\frac{{M_TK_T+K_RM_R}}{{N}}$ , which shows that 1) the one-shot sum-DoF scales linearly with the aggregate cache size in the network and 2) it is within a factor of $2$ to the information-theoretic optimum. Surprisingly, the identical and near optimal sum-DoF performance can be achieved using the hypercube approach with a much less file subpacketization.  
\end{abstract}
%--------------------------------------------------------------
\section{Introduction}
\label{section: intro}
Wireless traffic has grown dramatically in recent years due to the increasing mobile data demand, especially due to video delivery services \cite{cisco2016global}. One promising approach to %release the traffic pressure 
handle this traffic bottleneck is to exploit local cache memories at end user devices or network edge nodes (e.g., small cell base station) to pre-store part of the contents (e.g, movies) which might be requested in the near future. With the help of these cache nodes, the system can serve users with a much higher rate and lower latency \cite{maddah2014fundamental,wan2016caching,ji2016fundamental,
maddahali2015interference,Naderializadeh2017interference,Hachem2018interference,
bidokhti2017gaussian, Sengupta2017fog,shariatpanahi2017physical}. 
Among all schemes based on caching approaches, coded caching, introduced in \cite{maddah2014fundamental}, has attracted significant attention. In particular, Maddah-Ali and Niesen considered a {\em shared link network} and studied the problem of minimizing the worst-case traffic load or {\em rate}. It was shown that prefetching packets of the library files in a uniform manner during the placement phase, and employing coded scheme based on linear index code during the delivery phase, is sufficient to provide optimal rate under uncoded cache placement \cite{wan2016caching,yu2018uncoded}. Later, the idea of coded caching has been extended to Device-to-Device (D2D) caching networks \cite{ji2016fundamental}, %hierarchical caching networks \cite{Karamchandani16}, 
 multi-server caching networks \cite{Shariatpanahi2016multiserver} and combination caching networks \cite{wan2017survey}, where the channels between the transmitters and receivers are either wireline channel or broadcast noiseless channel.

The concept of coded caching was also extended to the wireless channels with the 
consideration of superpositions (e.g., interference) of transmitted signals  
\cite{maddahali2015interference,bidokhti2017gaussian,Naderializadeh2017interference,
Hachem2018interference,Sengupta2017fog,shariatpanahi2017physical}. 
In \cite{maddahali2015interference}, the authors considered a three-user interference channel where only transmitters are equipped with cache memories (no cache memories at the receivers) and showed that via a specific cache prefetching strategy, an efficient delivery scheme can be designed by exploiting the gains based on interference cancellation and interference alignment. {In  \cite{bidokhti2017gaussian}, the additive Gaussian channel in a broadcast setting with cache-aided receivers was studied.} Later, the study was extended to the case where both transmitters and receivers are equipped with cache. Moreover, cache-aided fog radio access network was also investigated in \cite{Sengupta2017fog}. 

As shown in above works, the remarkable multiplicative gain of coded caching in terms of network aggregate memory has been established in the asymptotic regime when the number of packets per file $F$ scales to infinity. It has been shown that in most of the cases, to achieve the desired caching gain, $F$ has to increase exponentially as a function of the number of nodes in the network. The finite length analysis of coded caching of shared link network was initiated in \cite{Shanmugam2016}, and was later characterized via combinatorial designs \cite{Shanmugam2017rs1}. 
The finite length analysis of coded caching in other network topologies other than shared link and MIMO broadcast channel is very limited. {In \cite{lampiris2018adding}, the authors considered a MISO broadcast channel with $L$ transmitting antennas and showed that reduced subpacketization can be achieved}. In addition, they extended the achievable scheme to the cache-aided interference networks. 
In \cite{woolsey2017towards}, we considered a D2D caching network over noiseless broadcast channel model and introduced a combinatorial design called {\em hypercube}, and the corresponding placement and coded delivery schemes with a substantially lower subpacketization level while still achieving order optimal throughput.  

In this paper, we consider the general wireless interference network with cache memories equipped at both the transmitter and receiver sides. In particular, we consider a wireless interference network with $K_T$ transmitters and $K_R$ receivers, each equipped with a local cache memory of size $M_T$ and $M_R$ files, from a library of $N$ files. We restrict the communication scheme to one-shot linear schemes due to its practicality. This network model was considered by Naderializadeh, Maddah-Ali and Avestimehr in \cite{Naderializadeh2017interference}. {We will demonstrate how to reduce the subpacketizaiton level (i.e., $F$) according to a deterministic combinatorial design called the hypercube approach}. %, which is the first attempt in the literature for this problem according to our knowledge.} 

Our main contribution in this paper is two-fold. First, based on the hypercube cache placement introduced in \cite{woolsey2017towards}, we designed a cache placement scheme at both transmitters and receivers, and proposed a linear one-shot delivery scheme by exploiting zero-forcing opportunities via transmitter collaboration and cache-induced interference cancellation opportunities at receivers' side. The proposed scheme achieves an order-wise subpacketization level reduction compared to that achieved in \cite{Naderializadeh2017interference}.  
Second, when $\frac{K_TM_T+K_RM_R}{N} \leq K_R$, the proposed scheme achieves a one-shot sum-DoF of $\frac{K_TM_T+K_RM_R}{N}$, which is within a factor of $2$ to the optimum as shown in \cite{Naderializadeh2017interference}. More importantly and surprisingly, it achieves the same sum-DoF as in \cite{Naderializadeh2017interference}. This implies that there is no any loss in terms of one-shot sum-DoF by using the proposed scheme while requiring a much less file subpacketization. In the rest of the paper, we will refer the scheme in \cite{Naderializadeh2017interference} as NMA scheme. %Due to the space limit, we will ignore the proofs of the theorems, which can be found in \cite{zhang2018hypercube}. 
    
\section{Network Model and Problem Formulation}
\label{sec: Network Model and Problem Formulation}
We use the following notation convention. Calligraphic symbols denote sets. We use $|\cdot|$ to represent the cardinality of a set or the length of a vector or the norm of a random variable; 
%$[a:b]:=\left\{ a,a+1,\ldots,b\right\}$ and 
$\mathbb{Z}^+$ denotes the integer set, ``$a\mod b$" denotes the module operation of $a$ modulo $b$, and $[n] := \{0, 1, \cdots, n-1\}$. We  also define the \textit{commutative product set} of two sets $\mathcal{A}$ and $\mathcal{B}$ as $\mathcal{A}\times \mathcal{B}=\left\{\{a,b\}:a\in \mathcal{A},b\in \mathcal{B}\right\}$.

\subsection{General Problem Formulation}
\begin{figure}  %\label{figure:1}
\centering
\includegraphics[width=0.38\textwidth]{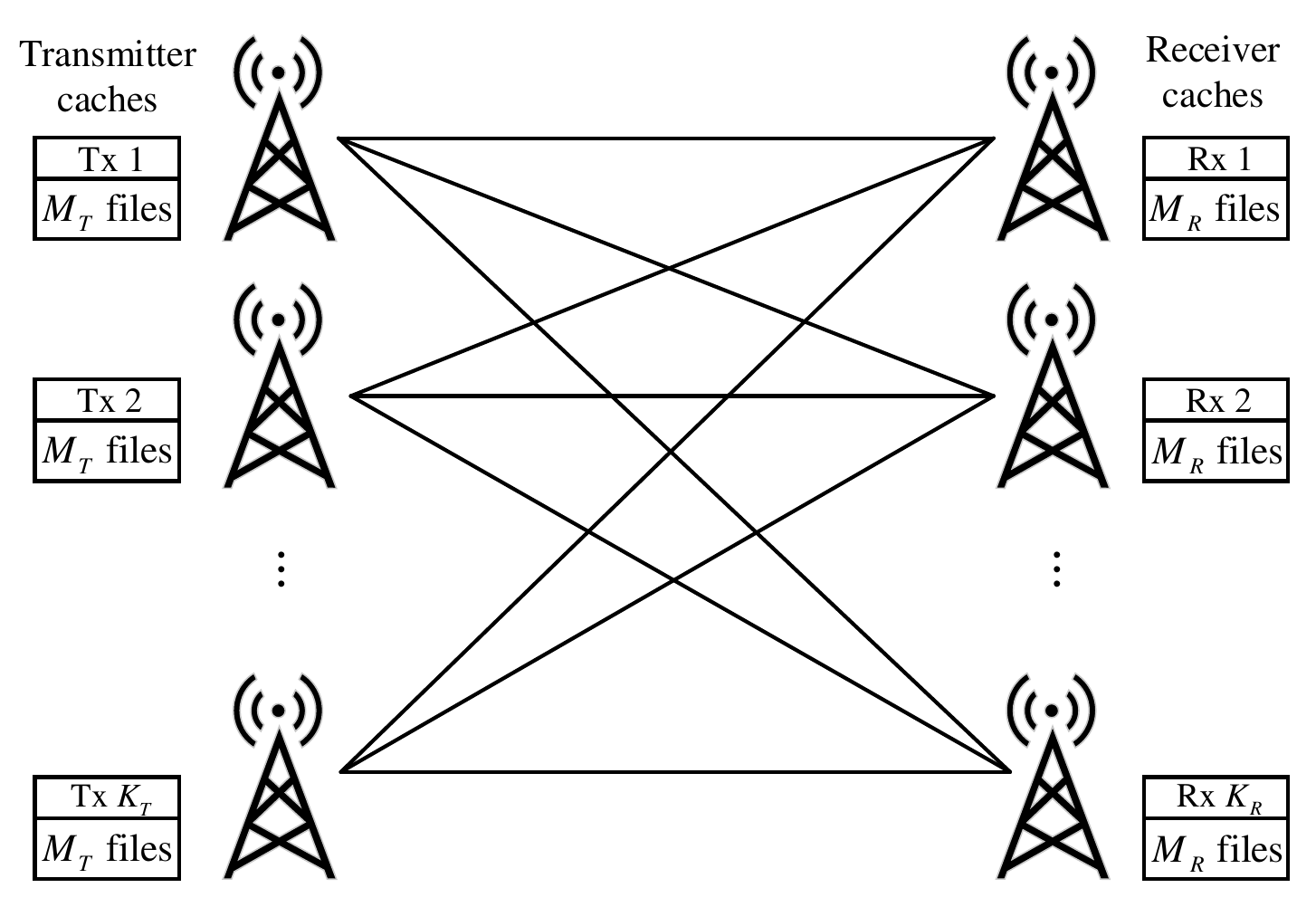}
\caption{Wireless interference network consisting of $K_T$ transmitters, each equipped with a cache of size $M_T$ files and $K_R$ receivers, each equipped with a cache of size $M_R$ files. The system also contains a library of $N$ files.}
\label{figure:1}
\vspace{-0.5cm}
\end{figure}

Consider a wireless interference network, as illustrated in Fig. \ref{figure:1}, which consists of $K_T$ transmitters and $K_R$ receivers, denoted by $\{\textrm{Tx}_i: i\in[K_T]\}$ and $\{\textrm{Rx}_j: j\in[K_R]\}$ respectively. The system contains a library of $N$ files denoted by $\{\mathcal{W}_n: n\in[N]\}$, where file $n$ contains $F$ packets $\{w_{n,p}: p\in[F]\}$ with %of 
size of $L$ bits each, i.e., $w_{n,p}\in \mathbb{F}_2^{L}$.\footnote{In this paper, we consider $L$ is a designed variable and equals to $|\Wc_n|/F$.} Transmitters and receivers are equipped with cache memories to store part of the file library. In particular, each transmitter and receiver are equipped with a local cache of size $M_T$ and $M_R$ files, respectively. %and each receiver is equipped with a local cache of size $M_R$ files. 
The communication channel between transmitters and receivers is modeled as discrete-time additive white Gaussian noise channel, which can be written as % at time slot $t$ is modeled as %by a discrete-time AWGN channel as
\begin{equation}
Y_j(t)=\sum_{i=0}^{K_T-1}h_{ji}S_i(t)+N_j(t),
\end{equation}
where $t$ is the index of the time slot. $S_i(t)\in \mathbb{C}$ is the complex transmit signal of Tx$_i$ at time slot $t$, satisfying the power constraint $\mathbb{E}[|S_i(t)|^2]\leq P$. $Y_j(t)$ is the received signal of Rx$_j$ and $N_j(t)\sim \mathcal{N}(0,1)$ is the complex additive white Gaussian noise (AWGN) %AWGN noise 
at receiver Rx$_j$. Moreover, $h_{ji}\in\mathbb{C}$ denotes the complex channel gain from Tx$_i$ to Rx$_j$, which is assumed to stay unchanged during the entire transmission process %whole course of communication 
and is known to all transmitters and receivers.  
The system operates in two phases: the \textit{prefetching phase} and the \textit{delivery phase} as described in \cite{Naderializadeh2017interference}. In the prefetching phase, each transmitter and receiver can store up to $M_TF$ and $M_RF$ arbitrary packets from the file library, respectively. %while each receiver can store up to $M_RF$ arbitrary packets from the file library. 
This phase is done without the prior knowledge of the receivers' requests. In the following delivery phase, the receivers' demands are revealed, i.e., each receiver Rx$_j$ will request a specific file $\mathcal{W}_{d_j},d_j\in[N]$ from the library. These requests are represented by a \emph{demand vector} defined as $\mathbf{d}\triangleq[d_0,\;d_1,\;\cdots,\;d_{K_R-1}]$. For a specific demand vector, since the receivers have already cached some packets of their requested files, the transmitters only need to deliver the remaining packets to those receivers. The task in this phase is to design the corresponding transmission procedure based on the cache placement in the prefetching phase so that the receivers' demands can be satisfied. In order to make sure that any possible demands can be satisfied, we require that the entire file library should be cached over all transmitters, %able to be ached at all transmitters, %union of all transmitters' cache memories should have at least a volume of $N$ files, 
i.e., $K_TM_T\geq N$.

For each cached packet $w_{n,p}\in \mathbb{F}_2^{L}$, the transmitter performs random Gaussian coding $\psi:\mathbb{F}_2^{L}\mapsto \mathbb{C}^{\hat{L}}$ to obtain the coded packet $\hat{w}_{n,p}\triangleq \psi(w_{n,p})$ which consists of $\hat{L}$ complex symbols. Assume that the communication will take place in $H$ blocks, each of which consists of $\hat{L}$ time slots. We also assume that one-shot linear scheme is employed in each block $m\in[1:H]$ to deliver a set of requested (coded) packets $\mathcal{P}_m$ to a subset of the receivers, denoted by $\mathcal{R}_m$. That is, each transmitter Tx$_i, i\in [K_T]$ will send a coded message 
\begin{equation}\label{linear combo coefficients}
s_i^m=\sum_{(n,p):w_{n,p}\in\mathcal{C}^{\rm T}_i\cap\mathcal{P}_m}\alpha_{i,n,p}^m\hat{w}_{n,p},
\end{equation} 
where $\mathcal{C}^{\rm T}_i$ denotes the cached contents of Tx$_i$ and $\alpha_{i,n,p}^m$ is the linear combination coefficients used by Tx$_i$ at the $m$-th block. Accordingly, the received signal of the intended receivers Rx$_j,j\in\mathcal{R}_m$ in the $m$-th block is
\begin{equation}
y_j^m=\sum_{i=0}^{K_T-1}h_{ji}s_i^m + n_j^m\,,
\end{equation}  
where $n_j^m\in\mathbb{C}^{\hat{L}}$ is the random %complex Gaussian 
noise at Rx$_j$ in block $m$. Each receiver will utilize its cached contents, consisting of packets stored in the prefetching phase, to subtract some of the interference caused by undesired packets. In particular, each receiver will perform a linear combination $\mathcal{L}_j^m(.)$ (if exists) to recover its requested packets from the overall received signals, as follows 
\begin{equation}
\mathcal{L}_j^m(y_j^m, \hat{\mathcal{C}}_j^{\rm R})=\hat{w}_{d_j,p} + n_j^m,
\end{equation} 
where $\hat{w}_{d_j,p}\in\mathcal{P}_m$ is the desired coded packet of Rx$_j$ and $\hat{\mathcal{C}}_j^{\rm R}$ denotes the Gaussian coded version of the packets cached by Rx$_j$. 

The one-shot linear sum-DoF %, introduced in \cite{Naderializadeh2017interference}, 
is defined as the maximum achievable one-shot linear sum-DoF for the worst case demands under a given caching realization \cite{Naderializadeh2017interference}, i.e.,
\begin{equation}
\mathsf{DoF}_{\rm L,sum}^{\left(\left\{\mathcal{C}_i^{\rm T}\right\}_{i=0}^{K_T-1},\left\{\mathcal{C}_j^{\rm R}\right\}_{j=0}^{K_R-1}\right)}=\inf\limits_{\mathbf{d}}\sup\limits_{H,\left\{\mathcal{P}_m\right\}_{m=1}^H}\frac{\left|\bigcup_{m=1}^H\mathcal{P}_m\right|}{H}.
\end{equation}
The one-shot linear sum-DoF of the network is correspondingly defined as the maximum achievable one-shot linear sum-DoF over all possible caching realizations, i.e.,
\begin{equation}
\begin{aligned}
&\mathsf{DoF}_{\rm L,sum}^{\ast}(N,M_T,M_R,K_T,K_R)=\\
&\qquad\qquad\quad\sup\limits_{\left\{\mathcal{C}_i^{\rm T}\right\}_{i=0}^{K_T-1},\left\{\mathcal{C}_j^{\rm R}\right\}_{j=0}^{K_R-1}}\mathsf{DoF}_{\rm L,sum}^{ \left(\left\{\mathcal{C}_i^{\rm T}\right\}_{i=0}^{K_T-1},\left\{\mathcal{C}_j^{\rm R}\right\}_{j=0}^{K_R-1}\right)},
\end{aligned}
\end{equation}
in which the cached contents of all transmitter and receivers satisfy the memory constraints, i.e., $|\mathcal{C}_i^{\rm T}|\leq M_TF,\forall i\in[K_T]$ and $|\mathcal{C}_j^{\rm R}|\leq M_RF,\forall j\in[K_R]$.

\subsection{Combinatorial Cache Placement Design}
In this paper, the combinatorial cache placement design based on {\em hypercube}, proposed in \cite{woolsey2017towards} 
to reduce the subpacketization level in wireless D2D networks, %address the file partition complexity problem in device-to-device caching networks, is employed 
is adopted in the prefetching phase.
We will show that this scheme can also significantly reduce the required number of packets per file in cache-aided interference networks %considered in this paper 
while using a very different delivery scheme. The details of hypercube cache placement \cite{woolsey2017towards} is described as follows. 

\subsubsection{Hypercube cache placement design for wireless D2D caching networks}

%\textbf{()}
%{\bf Hypercube cache placement.} 
%The hypercube cache placement was originally proposed for the wireless Device-to-Device (D2D) networks.
Consider a wireless D2D network %network 
comprising a library of $N$ files, each with $F$ packets, and $K$ users, each of which is equipped with a local cache of size $M$ files. Denote $t\triangleq \frac{KM}{N}\in\mathbb{Z}^+$ as the number of times that each file is cached among %the overall caches of 
all users. 
%\mj{[We may need an example to illustrate the hypercube and add a figure, probably, the three dimensional figure will suffice. I added a Powerpoint file on the Dropbox and you may want to use it.]}\xz{I have added a paragraph introducing the hypercube placement using the 3d example.} 
In the hypercube cache placement, each file $\mathcal{W}_n$ is partitioned into $\left(\frac{N}{M}\right)^t$ subfiles, i.e., $\mathcal{W}_n=\left\{\mathcal{W}_{n, (\ell_0,\ell_1,\cdots,\ell_{t-1})}: \,\ell_j\in[\frac{N}{M}],\,j\in[t]\right\}$. In the prefetching phase, each user $u\in[K]$ caches a set of subfiles $\left\{\mathcal{W}_{n,(\ell_0,\ell_1,\cdots,\ell_{t-1})}: \,n\in[N]\right\}$, where $\ell_j=u\mod \frac{N}{M}$, for $j=\lfloor u/\left(\frac{N}{M}\right)\rfloor$, and $\ell_i\in[\frac{N}{M}]$ for any $i\neq j$. As a result, each user will cache $(\frac{N}{M})^{t-1}$ packets from each file $\mathcal{W}_n$. %Since each subfile has a size of $\frac{F}{(\frac{N}{M})^t}$ packets, 
It can be verified that  the total number of packets cached by any user is equal to $N(\frac{N}{M})^{t-1} = N \frac{F}{N/M} = MF$, %$N(\frac{N}{M})^{t-1}\frac{F}{(\frac{N}{M})^t}=MF$, 
satisfying the memory constraint. Under the hypercube file partition method, each packet will represent a lattice point with coordinate $(\ell_0,\ell_1,\cdots,\ell_{t-1})$ in the $t$-dimensional hypercube, and $\frac{N}{M}$ is the number of lattice points on each edge. The hypercube cache placement is illustrated via the following example. 

\begin{figure}  %\label{figure:1}
\centering
\includegraphics[width=0.46\textwidth]{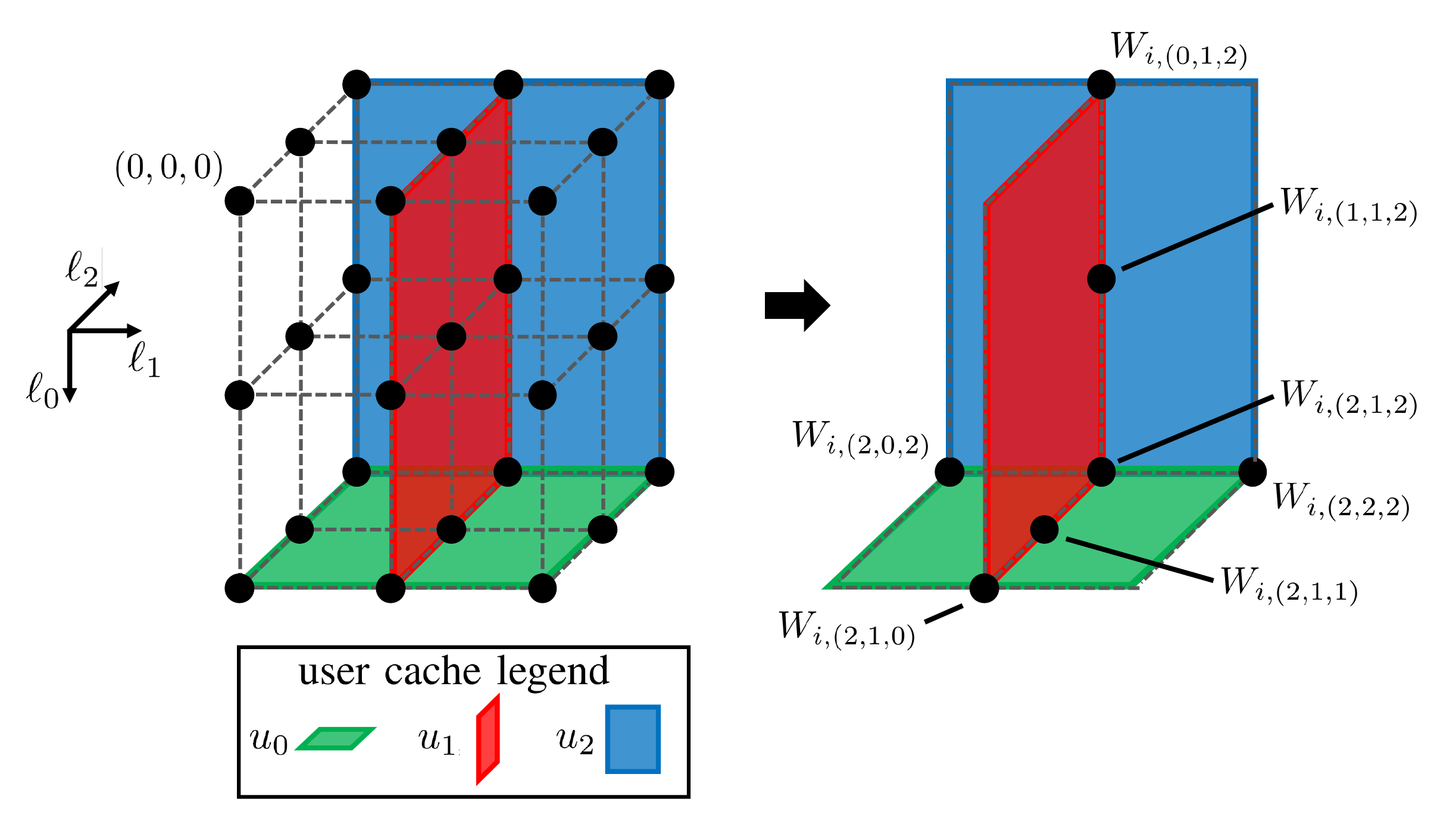}
\caption{A 3-dimensional example of the hypercube cache placement. Each subfile is represented by a unique lattice point in the 3-dimensional hypercube (cube). Each of the 9 users caches a set of packets represented by plane of lattice points. As a result, each user caches $9\times 9=81$ subfiles in total.}
\label{figure:2}
\vspace{-0.4cm}
\end{figure}

%\xz{3d hypercube example is added here.} 
\begin{example}
%We provide an example of the hypercube cache placement. 
Consider a set of 9 users labelled as $\{0,1,\cdots,8\}$ and a set of files $\Wc_n, n \in \{0,1,\cdots,8\}$. % $\mathcal{W}_i,i \in \{0,1,2,3\}$. 
We partition the users into $t\triangleq \frac{KM}{N}=3$ disjoint groups as $\mathcal{U}_0=\{0,1,2\}$, $\mathcal{U}_1=\{3,4,5\}$ and $\mathcal{U}_2=\{6,7,8\}$. We also split each file $\mathcal{W}_i$ into $3^3=27$ packets $\mathcal{W}_i=\{\mathcal{W}_{i,(\ell_0,\ell_1,\ell_2)}:\ell_0,\ell_1,\ell_2\in[3]\}$, each of which can be represented by a unique lattice point in the 3-dimensional cube (Fig. \ref{figure:2}). As a result, each lattice point will represent a set of $N=9$ packets, %4 subfiles, 
each from a distinct file. For the cache placement, each user caches all packets represented by a plane of lattice points of the cube. For example, user $u_0=2,u_1=4$ and $u_2=8$ will cache subfiles represented by the green, red and blue planes respectively. We can see that the set of packets $\{\mathcal{W}_{i,(2,1,2)}:i \in [9]\}$ represented by the lattice point $(2,1,2)$, which is intersection of the three orthogonal planes of different colors, is cached exclusively by users $u_0,u_1$ and $u_2$. Similarly, each subfile is cached by three distinct users.  
\hfill $\triangle$
\end{example}

\subsubsection{Hypercube cache placement design for cache-aided interference networks}

Different from the D2D settings in \cite{woolsey2017towards}, in cache-aided interference networks, we have explicit transmitters and receivers instead of D2D users. However, we can still group the transmitters and receivers into different {dimensions} 
%\mj{[We never explicitly say what a ``dimension" is.]}\xz{In the definition of the "hypercube permutation", I definded what a "dimension" is.} 
in order to implement the hypercube placement approach. Let $D_T\triangleq\frac{N}{M_T} \in \mathbb{Z}^+$ and $D_R\triangleq\frac{N}{M_R} \in \mathbb{Z}^+$. % and $D_R\triangleq\frac{N}{M_R}$, where we assume that $D_T$ and $D_R$ are both integers. 
For the set of $K_T=D_Tt_T$ transmitters $\{\textrm{Tx}_k: k\in[K_T]\}$, we denote the $t_T$ dimensions of the transmitters as $\mathcal{U}^{\rm T}_i=\{k:\lfloor \frac{k}{D_T} \rfloor=i\},\,i \in [t_T]$. Similarly, for the set of $K_R=D_Rt_R$ receivers $\{\textrm{Rx}_k: k\in[K_R]\}$, we denote the $t_R$ dimensions of the receivers as $\mathcal{U}^{\rm R}_j=\{k:\lfloor \frac{k}{D_R} \rfloor=j\},\, j \in [t_R]$. Here we require that $|\mathcal{U}^{\rm T}_i|=D_T,\,\forall i\in[t_T]$ and $|\mathcal{U}^{\rm R}_i|=D_R,\,\forall i\in[t_R]$.

The %corresponding general 
prefetching phase %and delivery phases are 
is described as follows.

\textit{Prefetching Phase}: The hypercube cache placement is employed at both the transmitters' and receivers' sides. That is, each file $\mathcal{W}_n$ is partitioned into ${D_T}^{t_T}{D_R}^{t_R}=(\frac{N}{M_T})^{t_T}(\frac{N}{M_R})^{t_R}$ disjoint equal-size subfiles,
%\footnote{We use the term ``subfiles" to represent a set of packets and the term ``packet" to represent the smallest unit in the file subpacketization.} 
denoted by
\begin{equation}
\mathcal{W}_n=\left\{\mathcal{W}_{n,\mathcal{T},\mathcal{R}}\right\}_{\substack{\mathcal{T}\in \mathcal{U}^{\rm T}_0\times\mathcal{U}^{\rm T}_1\times\cdots\times\mathcal{U}^{\rm T}_{t_T-1}\\
\mathcal{R}\in \mathcal{U}^{\rm R}_0\times\mathcal{U}^{\rm R}_1\times\cdots\times\mathcal{U}^{\rm R}_{t_R-1}}},
\end{equation}
where $\mathcal{T}=\{\tau_0,\tau_1,\cdots,\tau_{t_T-1}\}$ such that $\tau_i\in\mathcal{U}^{\rm T}_i, i\in[t_T]$ and $\mathcal{R}=\{r_0,r_1,\cdots,r_{t_R-1}\}$ such that $r_j\in\mathcal{U}^{\rm R}_j, j\in[t_R]$. Under this file partitioning strategy, each transmitter Tx$_i$ caches a set of subfiles $\{\mathcal{W}_{n,\mathcal{T},\mathcal{R}}:i\in\mathcal{T}\}$ and each receiver Tx$_j$ caches a set of subfiles $\{\mathcal{W}_{n,\mathcal{T},\mathcal{R}}:j\in\mathcal{R}\}$. As a result, the number of packets cached by Tx$_i$ is equal to 
\be
\label{eq: Tx memory}
N{D_T}^{t_T-1}{D_R}^{t_R}\frac{F}{{D_T}^{t_T}{D_R}^{t_R}}=M_TF\;\textrm{packets},
\ee
 %which satisfies the memory constraint of Tx$_i$. 
where $\frac{F}{{D_T}^{t_T}{D_R}^{t_R}}$ is the number of packets of each subfile and ${D_T}^{t_T-1}{D_R}^{t_R}$ is the number of subfiles cached by Tx$_i$. It can be easily verified that the memory constraint of transmitters is satisfied. Similarly, the number of packets cached by Rx$_j$ is equal to 
\be N{D_T}^{t_T}{D_R}^{t_R-1}\frac{F}{{D_T}^{t_T}{D_R}^{t_R}}=M_RF\;\textrm{packets},\ee 
which also satisfies the memory constraint. The hypercube cache placement approach in cache-aided interference networks is illustrated as follows. 

\begin{example}
\label{ex: prefetching interference}
Consider a wireless network with $K_T=4$ transmitters and $K_R=4$ receivers. Each transmitter and receiver is equipped with a cache of size $M_T=2$ and $M_R=2$ files respectively. The file library contains $N=4$ files denoted by $\mathcal{W}_0=A,\mathcal{W}_1=B,\mathcal{W}_2=C$ and $\mathcal{W}_3=D$.

%\textit{Prefetching Phase}: In this phase, 
In the prefetching phase, each file $\mathcal{W}_n$ is split into $2^2\times 2^2=16$ disjoint subfiles $\mathcal{W}_{n,\mathcal{T},\mathcal{R}}$ of equal sizes for any $\mathcal{T}\in\left\{\{0,2\},\{0,3\},\{1,2\},\{1,3\}\right\}$ and $\mathcal{R}\in\left\{\{0,2\},\{0,3\},\{1,2\},\{1,3\}\right\}$. Each subfile is then cached by the two transmitters in $\mathcal{T}$  and the two receivers in $\mathcal{R}$, respectively. For example, file $A$ is split into 16 subfiles:\footnote{{ With a slight abuse of notation, we write $A_{\{0, 2\},\{0, 2\}}$ as $A_{02,02}$ for simplicity and same for other symbols. }}
\begin{eqnarray}
&A_{02,02},\;A_{02,03},\;A_{02,12},\;A_{02,13},\nonumber\\
&A_{03,02},\;A_{03,03},\;A_{03,12},\;A_{03,13},\nonumber\\
&A_{12,02},\;A_{12,03},\;A_{12,12},\;A_{12,13},\nonumber\\
&A_{13,02},\;A_{13,03},\;A_{13,12},\;A_{13,13},\nonumber
\end{eqnarray}
where for example, $A_{02,02}$ is cached by transmitters Tx$_0$ and Tx$_2$ as well as receivers Rx$_0$ and Rx$_2$. The same file partitioning is done for files $B,C$ and $D$.

It can be seen that each transmitter will cache 8 subfiles of each file. Since each subfile contains $\frac{F}{16}$ packets, the total number of packets cached by each transmitter is $4\times 8\times\frac{F}{16}=2F$, which satisfies the memory constraint of the transmitters. Similarly, the memory constraint of the receivers is also satisfied. \hfill $\triangle$
\end{example}

\section{Main Results}
\label{section: main result}
The main results on the one-shot linear sum-DoF using the hypercube cache placement approach are presented in this section. Since for the case where $K_R<\frac{K_TM_T+K_RM_R}{N}$, it is argued in \cite{Naderializadeh2017interference} that the one-shot linear sum-DoF of $K_R$ is always achievable by utilizing only a fraction of the Tx and Rx cache memories, we focus on the case where $K_R \geq \frac{K_TM_T+K_RM_R}{N}$.
\begin{theorem}
\label{theorem:1}
For a $K_T\times K_R$ wireless interference network with a library of $N$ files, each consisting of $F$ packets, and with transmitter and receiver cache sizes of $M_TF$ and $M_RF$ packets respectively, given the hypercube cache placement approach employed in the prefetching phase, and for any $\delta\triangleq \frac{t_T}{t_R}\in\mathbb{Z}^+$, $D_R=\frac{K_R}{t_R}\geq \delta +1$, where $t_T\in[K_T],t_R\in[K_R]$, the one-shot linear sum-DoF of $\frac{K_TM_T+K_RM_R}{N}$ is achievable when $K_R \geq \frac{K_TM_T+K_RM_R}{N}$ with $F = (\frac{N}{M_T})^{t_T}(\frac{N}{M_R})^{t_R}\binom{D_R-2}{\delta-1}{\binom{D_R-1}{\delta}}^{t_R-1}\frac{(\delta!)^{t_R}}{\delta}(t_R-1)!$. 

\hfill  $\square$
\end{theorem}
 
\vspace{-0.3cm} 
\section{Achievable Delivery Scheme}\label{section: achievable scheme}
\begin{figure}
\centering
\includegraphics[width=0.5\textwidth]{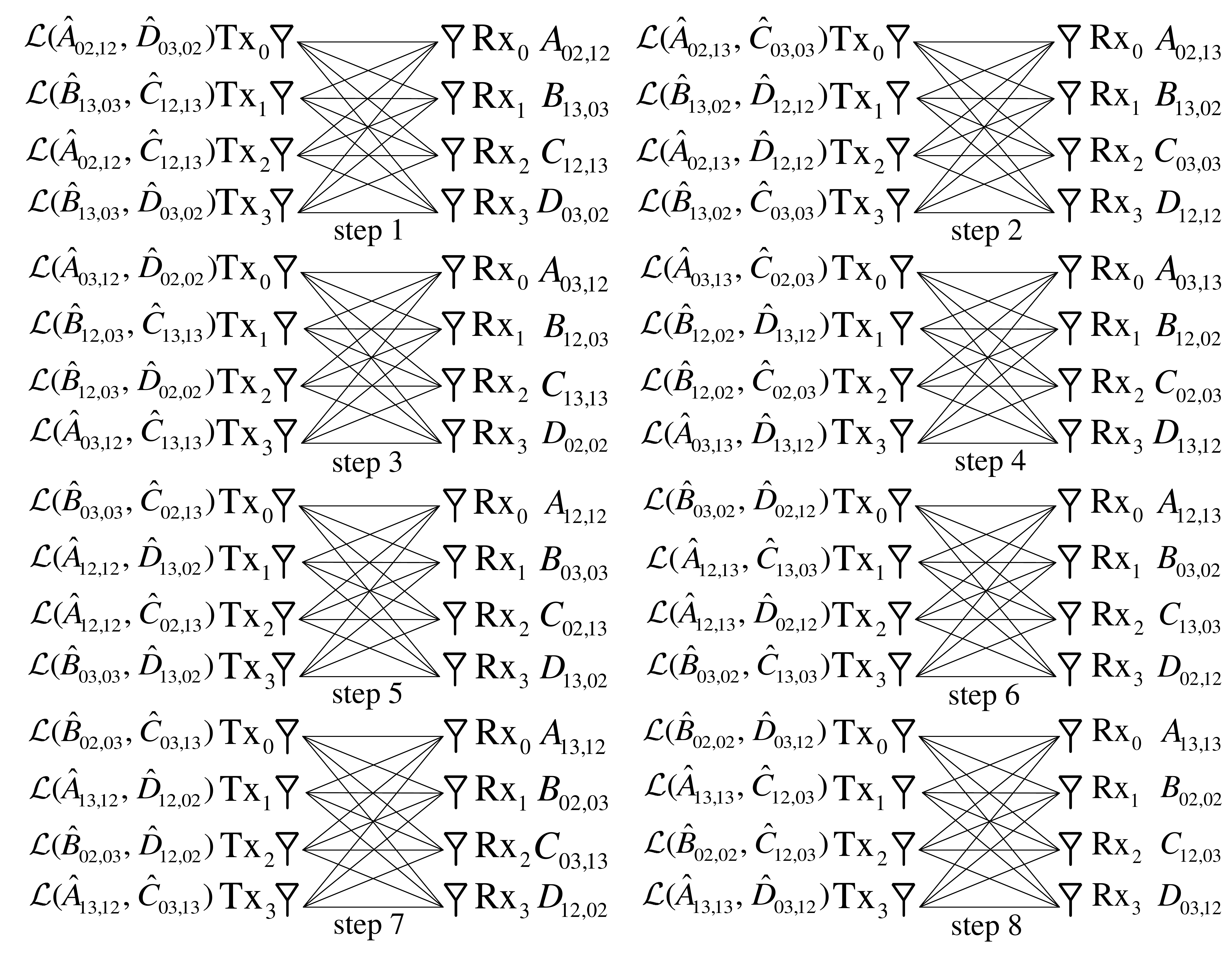}
\caption{Delivery phase for the Example \ref{ex: delivery interference}, %provided in Section \ref{subsection:A}, 
in which four receivers Rx$_j,j \in [4]$ request four different files $A,B,C$ and $D$ respectively. $\mathcal{L}({x,y})$ denotes some linear combination of $x$ and $y$, i.e., $\mathcal{L}(x,y)=\alpha x+\beta y$, where $\alpha$ and $\beta$ are some constants.} 
\label{figure:3}
\vspace{-0.3cm}
\end{figure}

We present the achievable delivery scheme under the hypercube cache placement via the following example. 
\begin{example}
\label{ex: delivery interference}
%\textit{Delivery Phase}: In this phase, 
We consider the same network settings as in Example \ref{ex: prefetching interference}. Let each receiver Rx$_j$ request a specific file $\mathcal{W}_{d_j}$ from the library. %WLOG, 
Without loss of generality,  we assume $\mathcal{W}_{d_0}=A,\mathcal{W}_{d_1}=B,\mathcal{W}_{d_2}=C$ and $\mathcal{W}_{d_3}=D$, respectively. In the prefetching phase, each receiver has already cached 8 subfiles of its requested file. Therefore, the transmitters only need to transmit the $16-8=8$ remaining subfiles to each receiver. More specifically, the following $32$ subfiles should be transmitted:
\[
 \begin{array}{ccc}
A_{02,12},\;A_{03,12},\;A_{12,12},\;A_{13,12},\\
A_{02,13},\;A_{03,13},\;A_{12,13},\;A_{13,13}\:\\
\end{array}
\Big\}\:\textrm{to Rx}_0 , 
\]
\vspace{-0.2cm}
\[
 \begin{array}{ccc}
B_{02,02},\;B_{03,02},\;B_{12,02},\;B_{13,02},\\
B_{02,03},\;B_{03,03},\;B_{12,03},\;B_{13,03}\:\\
\end{array}
\Big\}\:\textrm{to Rx}_1 , 
\]
\vspace{-0.2cm}
\[
 \begin{array}{ccc}
C_{02,03},\;C_{03,03},\;C_{12,03},\;C_{13,03},\\
C_{02,13},\;C_{03,13},\;C_{12,13},\;C_{13,13}\:\\
\end{array}
\Big\}\:\textrm{to Rx}_2  ,
\]
\vspace{-0.2cm}
\[
 \begin{array}{ccc}
D_{02,02},\;D_{03,02},\;D_{12,02},\;D_{13,02},\\
D_{02,12},\;D_{03,12},\;D_{12,12},\;D_{13,12}\:\\
\end{array}
\Big\}\:\textrm{to Rx}_3 . 
\]

In this example, since for $\delta=1$, $D_R=2$, and $t_R=2$, $\binom{D_R-2}{\delta-1}{\binom{D_R-1}{\delta}}^{t_R-1}=1$, no further file partitioning is needed. Hence, there are $32$ packets to be delivered to the receivers. 

We now show how the above $32$ packets can be grouped in $8$ subsets, each of which contains $4$ packets, such that the packets within the same subset can be transmitted simultaneously to the receivers without interference. Fig. \ref{figure:3} shows how the packet grouping and the corresponding transmissions are done. In each communication step, $4$ packets are delivered to the receivers simultaneously, and the interference between different users can be effectively eliminated by choosing proper linear combination coefficients at all transmitters. For example, in step 1 shown Fig. \ref{figure:3}, 4 packets $A_{02,12},B_{13,03},C_{12,13}$ and $D_{03,02}$ are transmitted to receivers Rx$_0$, Rx$_1$, Rx$_2$ and Rx$_3$ respectively. We write the transmit signals of each transmitter as a linear combination of some of these $4$ packets as follows:
\[
\begin{array}{ccc}
S_0=h_{32}\hat{A}_{02,12}-h_{13}\hat{D}_{03,02},\\
S_1=h_{23}\hat{B}_{13,03}-h_{02}\hat{C}_{12,13},\\
S_2=h_{01}\hat{C}_{12,13}-h_{30}\hat{A}_{02,12},\\
S_3=h_{10}\hat{D}_{03,02}-h_{21}\hat{B}_{13,03},
\end{array}
\]
where for each packet $\mathcal{W}_{n,\mathcal{T},\mathcal{R}}$, $\hat{W}_{n,\mathcal{T},\mathcal{R}}$ denotes its physical layer coded version. %, which contains a set of coded packets. 
As a result, after the interference cancellation over the air, the corresponding received signals by Rx$_0$, Rx$_1$, Rx$_2$ and Rx$_3$ are given by
%\[
\begin{align*}%{array}{ccc}
Y_0=&(h_{32}h_{00}-h_{30}h_{12})\hat{A}_{02,12}+(h_{23}h_{01}-h_{21}h_{03})\hat{B}_{13,03}\\
&+(h_{10}h_{03}-h_{13}h_{00})\hat{D}_{03,02}+N_0,\\
Y_1=&(h_{23}h_{11}-h_{21}h_{13})\hat{B}_{13,03}+(h_{32}h_{10}-h_{30}h_{12})\hat{A}_{02,12}\\
&+(h_{02}h_{11}-h_{01}h_{12})\hat{C}_{12,13}+N_1,\\
Y_2=&(h_{01}h_{22}-h_{02}h_{21})\hat{C}_{12,13}+(h_{32}h_{20}-h_{30}h_{22})\hat{A}_{02,12}\\
&+(h_{10}h_{23}-h_{13}h_{20})\hat{D}_{03,02}+N_2,\\
Y_3=&(h_{10}h_{33}-h_{13}h_{30})\hat{D}_{03,02}+(h_{23}h_{31}-h_{21}h_{33})\hat{B}_{13,03}\\
&+(h_{01}h_{32}-h_{02}h_{31})\hat{C}_{12,13}+N_3,
\end{align*}%{array}
%\]
where $N_i,i \in [4]$ is the random noise.

We can see that receiver Rx$_0$ can cancel the interference caused by %due to 
$B_{13,03}$ and $D_{03,02}$ 
since %$B_{13,03}$ and $D_{03,02}$ 
they have already been cached by Rx$_0$. Similarly, Rx$_1$, Rx$_2$ and Rx$_3$ can also cancel the interference %due to 
caused by undesired packets by utilizing their cached contents. Therefore, all the interference including inter-user interference and interference caused by cached packets %undesired subfiles, 
can be eliminated so that all receivers can decode their desired subfiles. It can be %easily 
verified that there exists such linear combinations and all receivers can decode their desired packets in all remaining $7$ communication steps. Hence, the $32$ packets, each consisting of $\frac{|\Wc_n|}{16}$ bits,  %, each consisting of $\frac{F}{16}$ packets,  
can be delivered to the receivers in 8 communication steps, each containing $\frac{F}{16} = 1$ resource block. As a result, a sum-DoF of %$\frac{32\times F/16}{8\times F/9}$$=
$\frac{32}{8} = 4=\frac{K_TM_T+K_RM_R}{N}$ can be achieved. 
%We can also verify that 
Hence, the proposed file subpacketization, placement, %in the prefetching phase, 
precoding and scheduling strategy in the delivery phase allow transmitters to collaboratively %to 
zero-force some of the outgoing interference and allow receivers to cancel the leftover interference %the known interference 
using cached contents for any receivers' demands.
\hfill $\triangle$
\end{example}

\subsection{Sketch of General Delivery Schemes}

In this section, we will sketch the general delivery scheme. %, whose detail is shown in \cite{zhang2018hypercube}. 
Let the receivers' demand vector be $\mathbf{d}=[d_0\;d_1\;\cdots\;d_{K_R-1}]$, %is revealed, 
i.e., receiver Rx$_j,j \in [K_R]$ requests the file $\mathcal{W}_{d_j}$. Since some subfiles of the requested file have already been cached by the receiver in the prefetching phase, the transmitters only need to send those subfiles which have not been cached by Rx$_j$, i.e., $\{\mathcal{W}_{d_j,\mathcal{T},\mathcal{R}}:j\notin\mathcal{R}\}$.  

Following similar approach in \cite{Naderializadeh2017interference}, we need to further partition the set of subfiles which need to be delivered to the receivers into smaller subfiles (or packets) so that they can be scheduled in %subsets 
groups of packets with size $t_T+t_R$ and delivered to the corresponding $t_T+t_R$ receivers simultaneously without interference. %interference-free. 
%More specifically, 
In particular, for each packet in a group of size $t_T+t_R$, %for any smaller subfile in the $t_T+t_R$ subset, 
it is desired by one particular receiver and can be cancelled by another $t_R$ receivers using %by utilizing 
their cached contents. In addition, some of the $t_T$ out of the $t_T+t_R$ transmitters can collaboratively zero-force the interference to another $t_T-1$ unintended receivers.  
To illustrate the further subpacketization of the subfiles, we need the following definition.
\begin{defn}
\textbf{(Hypercube Permutation)} Given a set of $D\times t$ points, denoted by $\mathcal{Q}$, we label each of these points by a unique number $u_{i,j}\in[Dt]$, where $i\in[t],j\in[D]$. Assume that these points can be divided into $t$ disjoint groups, which we refer to as \textit{dimensions}. Each dimension consists of $D$ points, denoted by $\mathcal{U}_i=\left\{u_{i,j}:\lfloor\frac{u_{i,j}}{D}\rfloor=i,j=0,1,\cdots,D-1\right\}$, $i\in[t]$. Define the \textit{hypercube permutation} of the set $\mathcal{Q}$, denoted by $\mathbf{\pi}^{\rm HCB}=[\,\pi(0)\;\pi(1)\;\cdots\;\pi(Dt-1)\,]$, as such a permutation of the $D\times t$ points that satisfies the following condition: For the set of users in $\mathcal{U}_i,i\in[t]$, the positions in the permutation (denoted by $pos(\cdot)$, meaning that $pos(u)=i$ if $\pi(i)=u$) of any two of them, $u_{i,j_1}$ and $u_{i,j_2}\,(j_1\neq j_2)$, should satisfy $|pos(u_{i,j_1})-pos(u_{i,j_2})|=kt,1\leq k \leq D-1,k\in \mathbb{Z^+}$ and $j_1,j_2\in[D]$. We also define the \textit{circular hypercube permutation} of a set $\mathcal{Q}$ as a way of arranging the elements of $\mathcal{Q}$ around a fixed table, and meanwhile, the corresponding arrangement should be a hypercube permutation.
\hfill $\lozenge$ 
\end{defn}
We illustrate the concept of \textit{hypercube permutation} and \textit{circular hypercube permutation} by the following example. 
\begin{example}
For $\mathcal{Q}=\{0,1,2,3\}$ with $t=2$ dimensions and $D=2$ points in each dimension, i.e., $\mathcal{U}_0=\{0,1\},\,\mathcal{U}_1=\{2,3\}$, we have 
\begin{eqnarray*}
\Pi^{\rm HCB}_{\mathcal{Q}}=\Big\{[\,0\;2\;1\;3\,],[\,0\;3\;1\;2\,],[\,1\;2\;0\;3\,],[\,1\;3\;0\;2\,],\\
\qquad[\,2\;1\;3\;0\,],[\,2\;0\;3\;1\,],[\,3\;1\;2\;0\,],[\,3\;0\;2\;1\,]\Big\}.
\end{eqnarray*}
It is clear that, for any two points within one dimension, $0,1\in\mathcal{U}_0$ or $2,3\in\mathcal{U}_1$, we have $|pos(0)-pos(1)|=|pos(2)-pos(3)|=2$, which satisfies the condition $|pos(u_{i,j_1})-pos(u_{i,j_2})|=t$ (note that $k=1$). Furthermore, we have $\Pi^{\rm HCB,circ}_{\mathcal{Q}}=\big\{[\;0\;2\;1\;3\;], [\;0\;3\;1\;2\;]\big\}$. 
\hfill $\triangle$
\end{example}

Based on the concept of circular hypercube permutations, we can illustrate the further subpacketization of the subfiles as follows. For any $j\in[K_R]$, $\mathcal{T}\in\mathcal{U}^{\rm T}_0\times\mathcal{U}^{\rm T}_1\times\cdots\times \mathcal{U}^{\rm T}_{t_T-1}$, and $\mathcal{R}\in\mathcal{U}^{\rm R}_0\times\mathcal{U}^{\rm R}_1\times\cdots\times\mathcal{U}^{\rm R}_{\lfloor\frac{j}{D_R}\rfloor}\setminus \{j\} \times\cdots \times\mathcal{U}^{\rm R}_{t_R-1}$, where $|\mathcal{T}|=t_T$ and $|\mathcal{R}|=t_R$, we partition $\mathcal{W}_{d_j,\mathcal{T},\mathcal{R}}$ into $\binom{D_R-2}{\delta-1}{\binom{D_R-1}{\delta}}^{t_R-1}\frac{(\delta!)^{t_R}}{\delta}(t_R-1)!$ packets. Using such
a further partition,, the delivery phase can be
carried out such that any receiver demand vector \textbf{d} can be
satisfied with a sum-DoF of $t_T + t_R$.

\section{Discussion}
In this section , we provide a comprehensive performance comparison of the hypercube cache placement based interference cancellation scheme and the NMA scheme. 

Each file in the library is partitioned into $(\frac{N}{M_T})^{t_T}(\frac{N}{M_R})^{t_R}$ subfiles in the hypercube based scheme while this number is $\binom{K_T}{t_T}\binom{K_R}{t_R}$ in the NMA scheme. In the delivery phase, to implement interference cancellation, each subfile which is requested by some receivers is further partitioned into $\Delta_{\rm HCB}\triangleq\binom{D_R-2}{\delta-1}{\binom{D_R-1}{\delta}}^{t_R-1}\frac{(\delta!)^{t_R}}{\delta}(t_R-1)!$ packets in the proposed hypercube based scheme and $\Delta_{\rm NMA}\triangleq t_R!\binom{K_R-t_R-1}{t_T-1}(t_T-1)!$ packets in the NMA scheme. Thus, the overall number of packets for a specific requested file required for these two schemes are $F_{\rm HCB} = {D_T}^{t_T}{D_R}^{t_R-1}(D_R-1)\Delta_{\rm HCB}$ and $F_{\rm NMA} = \binom{K_T}{t_T}\binom{K_R-1}{t_R}\Delta_{\rm NMA}$ respectively.  
\begin{figure}
\subfigure[$G(d,t,1)$ versus $t$]{
\includegraphics[width=0.235\textwidth]{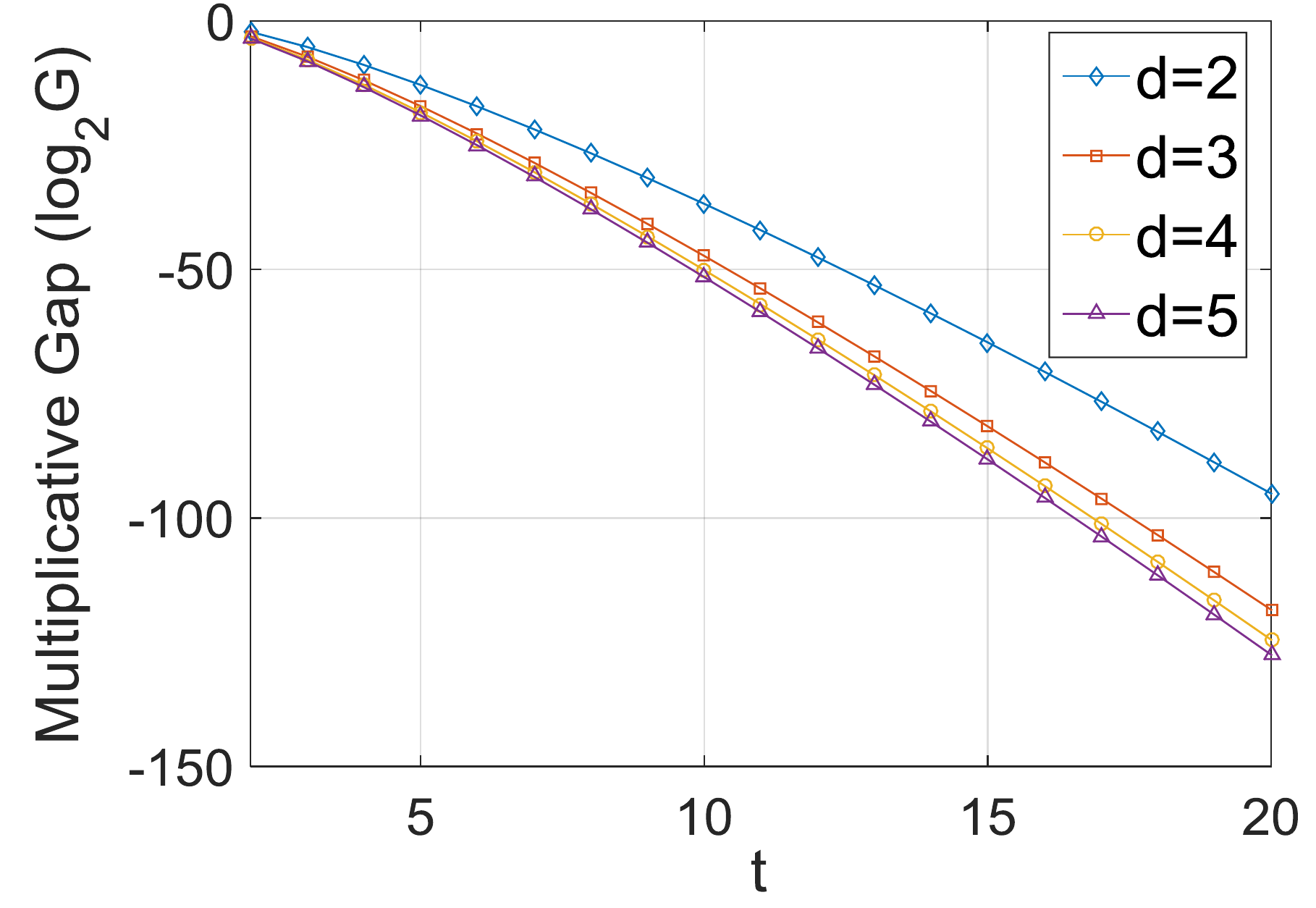}} %0.48
\subfigure[$G(d,t,1)$ versus $d$]{
%\vspace{-0.3cm}
\includegraphics[width=0.235\textwidth]{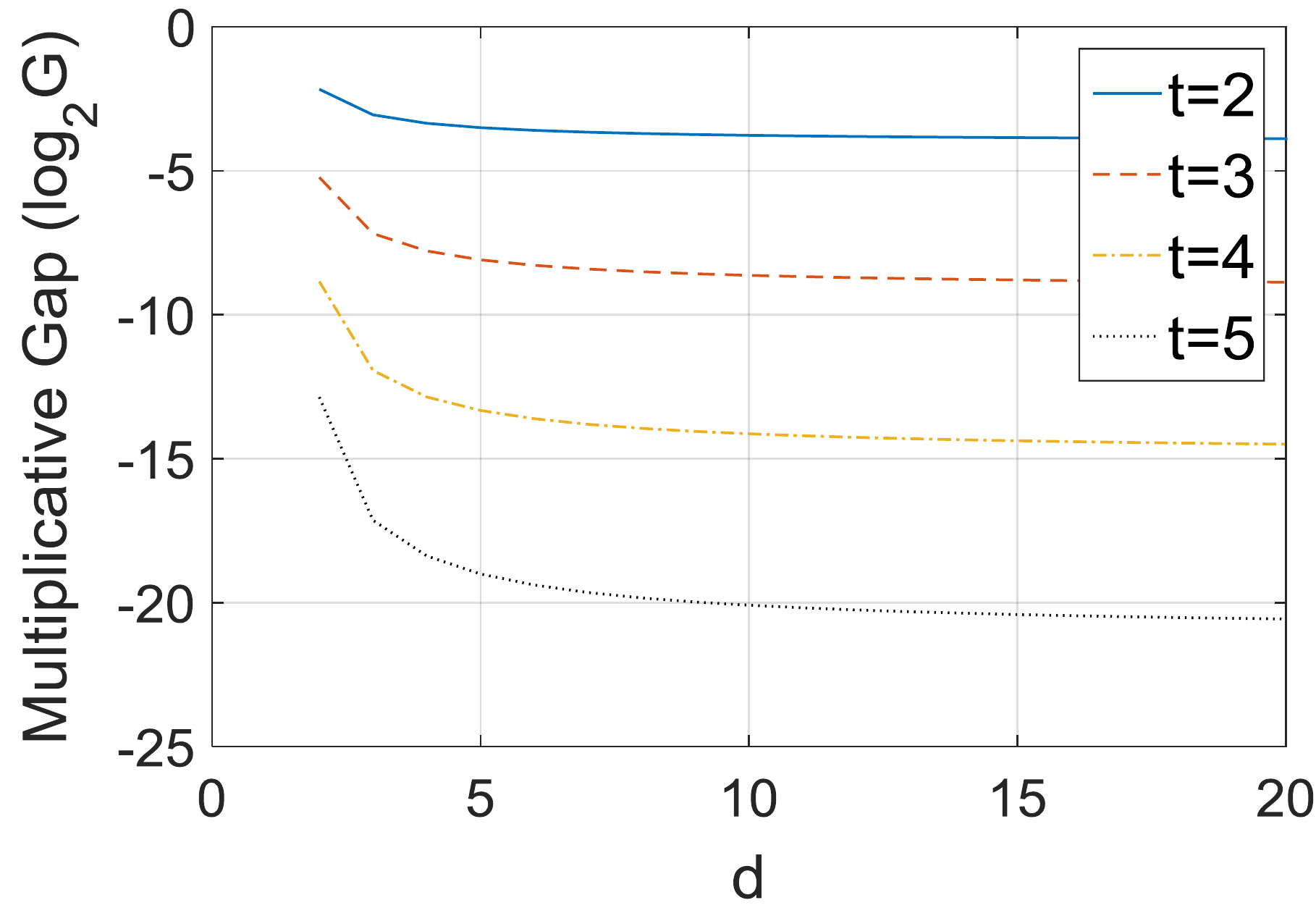}} %0.484
\caption{The multiplicative gap $G$ between the hypercube scheme and the NMA scheme. The comparison is down under the setting $t_T=t_R=t$, $N/M_T=N/M_R=d$, which implies $K_T=K_R=dt$. It can be seen that: (a) For a fixed $d$, $G$ decreases quickly as $t$ increases and approaches zero as $t$ goes to infinity, and (b) For a fixed $t$, $G$ converges to some specific non-zero value as $d$ goes to infinity.}
\label{figure:4}
\vspace{-0.3cm}
\end{figure} 
We define the \emph{multiplicative gap} of the subpacketization levels between these two schemes, denoted by $G$, as
$G\triangleq\frac{F_{\rm HCB}}{F_{\rm NMA}}$.
We next show that for any system parameters, the hypercube scheme has a strictly less subpacketization level than the NMA scheme.

%\textbf{Theorem 2:} 
\begin{theorem}
\label{theorem:2}For any system parameters $K_T,K_R,M_T,M_R$ and $N$ satisfying $t_T=\frac{K_TM_T}{N}\in \mathbb{Z}^+,t_R=\frac{K_RM_R}{N}\in \mathbb{Z}^+,D_T=\frac{K_T}{t_T}\in\mathbb{Z}^+,D_R=\frac{K_R}{t_R}\in\mathbb{Z}^+$ and $\delta\triangleq\frac{t_T}{t_R}\in\mathbb{Z}^+,D_{R}\ge \delta +1$, the  multiplicative gap $G$ is strictly less than $1$. Moreover, under the setting $t_T=\delta t_R=\delta t,D_T=D_R=d$, we have $\lim_{t\to\infty}G(d,t,\delta)=0$. More specifically, for fixed  $d,\delta$ and large enough $t$, we have 
\be 
G(d,t,\delta)\le \frac{C_0 e^{-(\delta+1)t}}{C_1^{(\delta-1)t}}
\ee
in which the constants are $C_0=\frac{2\pi \sqrt{\delta}(d-2)!e^{\frac{1}{d}+\frac{\delta}{d-1}}}{(d-\delta -1)!}$, $C_1=\frac{(d-1)!}{(d-\delta -1)!}$. For fixed $t$ and $\delta$, it holds that 
\begin{eqnarray}
\lim_{d\to\infty}G(d,t,\delta)&=&\frac{(t-1)!(\delta t)!}{\delta^{\delta t}t^{(2\delta+1)t-1}}. \label{eq: G d}
\end{eqnarray}
\hfill  $\square$
\end{theorem}

One important implication of Theorem 2 is the subpacketzation level reduction of the hypercube scheme over the NMA scheme, which yields a significant advantage since it holds for any possible system parameters while preserving the same one-shot linear sum-DoF gain. Fig. \ref{figure:4} shows the multiplicative gain $G(d,t,1)$ under logarithmic scale for the case when $\delta\triangleq \frac{t_T}{t_R}=1,t_T=t_R=t$ and $D_R=D_R=d$. It can be seen that the gap decreases exponentially as $t$ increases and goes to zero as $t$ goes to infinity (see Fig. \ref{figure:4}(a)) and $G$ converges to some specific value as $d$ goes to infinity (see Fig. \ref{figure:4}(b)).

\section*{Acknowledgments}
This work is supported by NSF~1817154  and NSF~1824558. 

\bibliographystyle{IEEEbib}
\bibliography{references_d2d}

\begin{thebibliography}{10}

\bibitem{cisco2016global}
Visual Networking~Index Cisco,
\newblock ``Global mobile data traffic forecast update, 2015--2020 white
  paper,''
\newblock {\em Document ID}, vol. 958959758, 2016.

\bibitem{maddah2014fundamental}
M.~A. Maddah-Ali and U.~Niesen,
\newblock ``Fundamental limits of caching,''
\newblock {\em Information Theory, IEEE Transactions on}, vol. 60, no. 5, pp.
  2856--2867, 2014.

\bibitem{wan2016caching}
K.~Wan, D.~Tuninetti, and P.~Piantanida,
\newblock ``On caching with more users than files,''
\newblock in {\em 2016 IEEE International Symposium on Information Theory
  (ISIT)}, July 2016, pp. 135--139.

\bibitem{ji2016fundamental}
M.~Ji, G.~Caire, and A.~F. Molisch,
\newblock ``Fundamental limits of caching in wireless d2d networks,''
\newblock {\em IEEE Transactions on Information Theory}, vol. 62, no. 2, pp.
  849--869, Feb 2016.

\bibitem{maddahali2015interference}
M.~A. Maddah-Ali and U.~Niesen,
\newblock ``Cache-aided interference channels,''
\newblock in {\em 2015 IEEE International Symposium on Information Theory
  (ISIT)}, June 2015, pp. 809--813.

\bibitem{Naderializadeh2017interference}
N.~Naderializadeh, M.~A. Maddah-Ali, and A.~S. Avestimehr,
\newblock ``Fundamental limits of cache-aided interference management,''
\newblock {\em IEEE Transactions on Information Theory}, vol. 63, no. 5, pp.
  3092--3107, May 2017.

\bibitem{Hachem2018interference}
J.~Hachem, U.~Niesen, and S.~N. Diggavi,
\newblock ``Degrees of freedom of cache-aided wireless interference networks,''
\newblock {\em IEEE Transactions on Information Theory}, vol. 64, no. 7, pp.
  5359--5380, July 2018.

\bibitem{bidokhti2017gaussian}
S.~S. Bidokhti, M.~Wigger, and A.~Yener,
\newblock ``Gaussian broadcast channels with receiver cache assignment,''
\newblock in {\em 2017 IEEE International Conference on Communications (ICC)}.
  IEEE, 2017, pp. 1--6.

\bibitem{Sengupta2017fog}
A.~Sengupta, R.~Tandon, and O.~Simeone,
\newblock ``Fog-aided wireless networks for content delivery: Fundamental
  latency tradeoffs,''
\newblock {\em IEEE Transactions on Information Theory}, vol. 63, no. 10, pp.
  6650--6678, Oct 2017.

\bibitem{shariatpanahi2017physical}
S.~P. Shariatpanahi, G.~Caire, and B.~H. Khalaj,
\newblock ``Physical-layer schemes for wireless coded caching,''
\newblock {\em arXiv preprint arXiv:1711.05969}, 2017.

\bibitem{yu2018uncoded}
Q.~Yu, M.~A. Maddah-Ali, and A.~S. Avestimehr,
\newblock ``The exact rate-memory tradeoff for caching with uncoded
  prefetching,''
\newblock {\em IEEE Transactions on Information Theory}, vol. 64, no. 2, pp.
  1281--1296, Feb 2018.

\bibitem{Shariatpanahi2016multiserver}
S.~P. Shariatpanahi, S.~A. Motahari, and B.~H. Khalaj,
\newblock ``Multi-server coded caching,''
\newblock {\em IEEE Transactions on Information Theory}, vol. 62, no. 12, pp.
  7253--7271, Dec 2016.

\bibitem{wan2017survey}
K.~Wan, D.~Tuninetti, M.~Ji, and P.~Piantanida,
\newblock ``State-of-the-art in cache-aided combination networks,''
\newblock in {\em 2017 51st Asilomar Conference on Signals, Systems, and
  Computers}, Oct 2017, pp. 641--645.

\bibitem{Shanmugam2016}
K.~Shanmugam, M.~Ji, A.~M. Tulino, J.~Llorca, and A.~G. Dimakis.,
\newblock ``Finite-length analysis of caching-aided coded multicasting,''
\newblock {\em IEEE Transactions on Information Theory}, vol. 62, no. 10, pp.
  5524--5537, Oct 2016.

\bibitem{Shanmugam2017rs1}
K.~Shanmugam, A.~M. Tulino, and A.~G. Dimakis,
\newblock ``Coded caching with linear subpacketization is possible using
  ruzsa-szemeredi graphs,''
\newblock in {\em 2017 IEEE International Symposium on Information Theory
  (ISIT)}, June 2017, pp. 1237--1241.

\bibitem{lampiris2018adding}
E.~Lampiris and P.~Elia,
\newblock ``Adding transmitters dramatically boosts coded-caching gains for
  finite file sizes,''
\newblock {\em IEEE Journal on Selected Areas in Communications}, vol. 36, no.
  6, pp. 1176--1188, 2018.

\bibitem{woolsey2017towards}
N.~Woolsey, R.-R. Chen, and M.~Ji,
\newblock ``Towards practical file packetizations in wireless device-to-device
  caching networks,''
\newblock {\em arXiv preprint arXiv:1712.07221}, 2017.

\end{thebibliography}

\end{document}